\documentclass[12pt,preprint]{aastex}

\shorttitle{Fine structures in type IIIb bursts}
\shortauthors{Wang}

\begin{document}


\title{A Scenario for the Fine Structures of Solar Type IIIb Radio
       Bursts Based on the Electron Cyclotron Maser Emission}


\author{C. B. Wang$^{1, 2, 3}$}
\affil{$^1$CAS Key Laboratory of Geospace Environment,
   School of Earth and Space Science, University of Science and Technology
   of China, Hefei 230026, Anhui, China}

\affil{$^2$Collaborative Innovation Center of Astronautical Science and Technology, China}

\affil{$^3$Mengcheng National Geophysical Observatory, School of Earth and Space Science, University of Science and
Technology of China, Hefei 230026, Anhui, China}

\email{cbwang@ustc.edu.cn}




\begin{abstract}
A scenario based on the electron cyclotron maser emission is proposed for the fine structures of solar radio emission in
the present discussion. It is suggested that under certain conditions modulation of the ratio between the plasma
frequency and electron gyro-frequency by ultra low frequency waves, which is a key parameter for excitation of the
electron cyclotron maser instability, may lead to the intermittent emission of radio waves. As an example, the
explanation of the observed fine-structure components in the solar type IIIb burst is discussed in detail. Three primary
issues of the type IIIb bursts are addressed: 1) what is the physical mechanism that results in the intermittent emission
elements that form a chain in the dynamic spectrum of type IIIb bursts, 2) what causes the split pair (or double stria)
and the triple stria, 3) why in the events of fundamental-harmonic pair emission there is only IIIb-III, but IIIb-IIIb or
III-IIIb cases are very rarely observed.
\end{abstract}


\keywords{Solar Radio Bursts; Fine Strucutres; Type IIIb Bursts; Electron Cyclotron Maser Emission}



\section{Introduction}

There are a rich variety of fine structures of solar radio emission in the form of wide-band pulsations in emission and
absorption with different periods, rapid bursts, narrow-band patches \citep{Chernov2011}. Specially, the type IIIb solar
radio burst is a chain of several elementary bursts which appear in dynamic spectra as either single, double, or triple
narrow-banded striations, but their envelope resembles a type III bursts \citep{Noe72,Noe75,Dulk80}. The study of these
fine structures is a key to understanding and verifying of different emission mechanism of solar radio bursts.

Type III bursts are obviously the most frequent bursts observed from the solar corona. It is generally believed that they
are generated by a beam of fast electrons moving along the corona or interplanetary magnetic field line. The plasma
emission mechanism is now the most commonly accepted standard model for the type III burst generation, which was first
described by \citet{Ginzburg58} and then refined by a number of researchers \citep{Sturrock64,ZheleznyakovZ70,Smith76,
Melrose80,Goldman83,Melrose85,Robinson94,Robinson98a,Robinson98b,Robinson98c,Li2008}. In the plasma-emission theory, it
is suggested that streaming electrons first excite Langmuir waves, and then part of the energy of the enhanced Langmuir
waves is converted into electromagnetic (EM) emissions with frequencies close to the local plasma frequency and its
harmonic through non-linear wave-wave interaction. The differences in existing works are mainly in the details by which
the excited Langmuir waves are partly converted to EM waves. In addition, Huang (1998) suggested that Langmuir waves can
also be converted directly into EM waves.

Alternatively, a direct emission mechanism based on the electron cyclotron maser (ECM) emission mechanism has been
proposed for type III solar radio bursts \citep{Wu2002,Wu2004,Wu2005,Yoon2002}. Two basic assumptions are made in this
ECM emission model. First, it is postulated that density-depleted magnetic flux tubes exist in the corona, because in the
low-beta corona a small perturbation of the ambient magnetic field in a flux tube can result in considerable density
change due to pressure balance \citep{Duncan79,Wu2006}. Second, flare-associated streaming electrons possess a velocity
distribution with a perpendicular population inversion, which is unstable to ECM instability so that the ordinary
(O)-mode or extraordinary (X)-mode EM waves can be directly amplified with frequency near the electron gyro-frequency or
its harmonics in the density-depleted tube. The propagation of amplified radio emission is initially confined within the
magnetic flux tube until it arrives at a point where the local exterior cutoff frequency is equal to the exciting wave
frequency. On the basis of this model, a number of long-standing issues raised from observation can be resolved simply
and self-consistently, such as the existence of fundamental-harmonic (F-H) pairs and "structureless" bursts with no
visible F-H structure, the same apparent source positions of F- and H-band waves observed at a fixed frequency
\citep{McLean71,Stewart72,Stewart74,Dulk80}, the F component being more directive than H component of F-H pairs
\citep{Caroubalos74,Dulk80,Suzuki85}, the starting frequency of F waves often being just about one-third to one-fifth
that of the H components \citep{Dulk80}. However, more discussion is needed. An outstanding theoretical issue to be
addressed, is to explain the physical origin of the fine-structure components that occasionally appear in the dynamic
spectrum of type IIIb radio emission.

Type IIIb bursts often consist of a large number of stria bursts but their envelope resembles a type III burst. Early
observational results concerning type IIIb bursts are reported in \citet{Ellis67}, \citet{Ellis69}, \citet{Noe72},
\citet{Noe75}, \citet{Dulk80}. In general, there are several types of fine structures such as double stria burst, triple
stria burst, and fork burst, etc. The spectral properties of type IIIb bursts are discussed and reviewed in
\citet{McLean85} and \citet{Suzuki85}. Although type IIIb radio bursts represents merely a fraction of type III emission
events, theoretical study of such a phenomenon is essential. In principle, any acceptable theory of type III emission
should be able to explain type IIIb bursts in the same context.

The common belief in the plasma emission mechanism is that density inhomogeneities in the background plasma create a
clumpy distribution of Langmuir waves and are the cause of the type IIIb fine structure
\citep{Takakura75,Melrose83,Reid2014}. On the other hand, \citet{Li2011a,Li2011b} proposed that localized disturbances in
the electron-temperature and/or ion-temperature in the corona may also be responsible for the stria and the type IIIb
bursts. However, the numerical simulation results show that the fine structures are more pronounced for H emission than
the F emission \citep{Li2012}. This is inconsistent with the interpretations of many observers that stria bursts occur
more often in the F component than in the H component in the case of F-H pair burst.

Based on the ECM emission scenario, \citet{Zhao2013} recently considered that a linearly polarized homochromous Alfv\'en
wave can modulate ECM emission process and may be responsible for the fine structure of type IIIb bursts. In this
article, we will propose another scenario for the fine structures of solar radio emission. While we also suggest that
under certain conditions ECM emission modulation by ultra-low frequency waves may lead to the intermittent emission of
radio waves, the physical process for the modulation is basically different from that suggested by \citet{Zhao2013}. The
present scenario is preliminary, but it explains most of the observed features.

The organization of the discussion is as follows. In Section 2 we review briefly the essence of the ECM emission
mechanism for type III bursts that is the basis of the subsequent discussion. Then, in Section 3 a physical model is
suggested for type IIIb emission. Discussion and conclusions are presented in Section 4 and Section 5.

\section{Electron-Cyclotron Maser Emission of Type III Burst}


As mentioned in the Introduction, the plasma emission theory is now the most widely known scenario for solar type III
bursts. The ECM emission scenario may be not familiar to some readers, so before discussing the fine structures of solar
type IIIb bursts, it is better to describe briefly the essential points of ECM emission scenario for normal type III
bursts without fine structures \citep{Wu2002,Wu2005,Yoon2002}. The motivation of the ECM emission scenario is mainly
coming from two observations. 1) Type III bursts are generated by fast electrons associated with solar flares. Most solar
flares occur in active regions where the magnetic field is stronger than that elsewhere at the same altitude. Effect of
the ambient magnetic field may be important for the emission process. 2) Observations find that waves of the F component
and the H component of type III bursts with same frequency have coincidental source regions
\citep{McLean71,Stewart72,Stewart74,Dulk80}. This implies that the radiation of type III bursts may produced in a
density-depleted flux tube as being noted firstly by \citet{Duncan79}.

Figure 1 is a graphic scheme that summarizes one possible ECM emission scenario for solar type III bursts. We assume that
a solar flare occurs somewhere above an active region through the magnetic reconnection between a magnetic loop and an
open magnetic flux tube in the corona. Energetic electrons are generated during the impulsive phase of the flare. In
general, these energetic particles may occur either inside or outside the density-depleted flux tube. We are only
interested in those that occur inside an open flux tube with plasma density that is low enough to ensure the ECM emission
is workable. Meanwhile, the energetic electrons may move away from the reconnection site along open field lines in both
upward and downward directions. In the present discussion we are interested in the upward streaming electrons. It is
believed that enhanced turbulent Alfv\'en waves exist pervasively in the solar corona. These waves can pitch-angle
scatter the streaming electrons. The pitch-angle scattering accelerates the electrons in the transverse direction, and is
more effective for fast electrons. As a result, the streaming electrons deform into a crescent-shaped distribution
\citep{Wu2002,Wu2012} with a perpendicular population inversion, which is unstable to the ECM instability so that the
O-mode or X-mode EM waves can be directly amplified with frequency near the electron gyro-frequency or its harmonics in
the density-depleted tube. The cutoff frequency of either the X-mode or the O-mode inside the tube is significantly lower
than that outside the tube. The true source region, where the radio wave is generated, is inside the tube, and the wave
is confined in the tube during propagation if the wave frequency is below the exterior cutoff frequency (thus, the true
source region is not observable). This wave cannot leave the tube until it reached an altitude where the local exterior
cutoff frequency becomes lower than the wave frequency. We name the region where the wave leaves the density-depleted
flux tube as apparent source region. We consider that the observed source regions of F and H waves are actually apparent
regions rather than the true source regions. Hence, waves, regardless the locations of their generation, with the same
frequency would exit at the same altitude. In addition, We would like to reiterate that figure 1 is just for illustration
purpose. There are a number of solar flare models in the literatures \citep{Benz2008}.

The details of the ECM instability, the generation mechanism of
radio emission suggested above, have already been given in several
articles \citep{Wu2002,Yoon2002,Wu2005,Chen2005}, so we will not
repeat the discussion here. In principle, both X-mode wave and
O-mode wave can be amplified by the maser instability. Because the
O-mode waves have much lower level of spontaneous emission, the
emission level of the O-mode waves may be insignificant in
comparison with that for X-mode waves as discussed in
\citet{Wu2002} and \citet{Yoon2002}. In the following, we will
mainly pay attention to the X-mode waves. In other words, we
assume that the X1 mode waves and X2 mode waves with frequency
near the electron gyro-frequency and its harmonics are
corresponding for the observed F components and H components of
type III bursts, respectively. Nevertheless, the discussion is
also applicable for the O-mode waves.

The ratio $\omega_{pe}/\Omega_e$  ($\omega_{pe}$ is the electron plasma frequency and $\Omega_e$ is the electron gyro
frequency) is a crucial parameter for the maser instability. Here we present figure 2 in which the maximum growth rate is
plotted versus the ratio $\omega_{pe}/\Omega_e$  for the X1 and X2 waves. It is seen from figure 2 that the X1 waves are
amplified by the streaming electrons only if the ratio $\omega_{pe}/\Omega_e$ falls within the range $0.1<
\omega_{pe}/\Omega_e<0.4$. Once the ratio $\omega_{pe}/\Omega_e$ is beyond the upper limit, emission of X1 waves is
considerably weakened. However, on the other hand, the condition for X2 waves is $0.1< \omega_{pe}/\Omega_e<1.4$, so that
the range is much broader. The main reason is that X2 waves have frequencies far above the cutoff frequency, and have
frequencies near the second harmonic of the gyro frequency.

On the basis of this model, several important issues for type III bursts raised from observation can be understood
naturally. For examples:

$\bullet$ Whether we observe F-H pairs or "structureless" bursts without F-H pairs depends on the frequency ratio
$\omega_{pe}/\Omega_e$ in the flux tube. If the ratio $\omega_{pe}/\Omega_e$  falls within the range
$0.1<\omega_{pe}/\Omega_e<0.4$, the burst would be F-H pairs. On the other hand, if $\omega_{pe}/\Omega_e$ falls within
the range $0.4< \omega_{pe}/\Omega_e<1.4$, the "structureless" burst would be observed, and it is the H component.

$\bullet$ The observed source region is the apparent region whose altitude is dependent only on the wave frequency.
Whether it is F component or H component, waves with the same frequency would exit at the same altitude. Thus, the same
apparent source positions of F- and H- band waves would be observed at a fixed frequency.

$\bullet$ The F waves initially propagate in the oblique direction, while H waves are excited with wave-vectors primarily
along nearly perpendicular direction. Moreover, before escaping from the density-depleted tube, F waves must propagate
much longer along the duct than H waves that were generated at the same true source region. Hence, one would expect that
F waves are generally more directive than H waves as one can see clearly from the ray-trace results shown by the Figure 8
in the paper by \citet{Yoon2002}.

$\bullet$ The emission of F waves starts at an altitude higher than that for H waves, because the exciting of F waves
requires a sufficiently large beam momentum (Wu et al., 2005). Hence, one would expect that the ratio of the starting
frequencies of H components to those of the F components is generally higher than two \citep{Dulk80}.

\section{An Interpretation of type IIIb emission}

\subsection{Basic consideration}
The most conspicuous feature of type IIIb emission is that in cases of F-H pair emission the fine structure only occurs
in the F component \citep{Noe72,Suzuki85}. This outstanding feature implies that in type IIIb bursts the F component is
intermittently suppressed. The key question is what causes the suppression.

As remarked before, the ratio $\omega_{pe}/\Omega_e$ plays a crucial role in the ECM instability. The amplification of
the X1 mode requires $0.1< \omega_{pe}/\Omega_e<0.4$. Once the ratio is beyond its upper limit, emission of X1 waves is
suppressed. On the other hand, the condition for X2 waves is  $0.1< \omega_{pe}/\Omega_e<1.4$, so that the range is much
broader. Evidently, if the density and/or magnetic field vary spatially in a quasi-periodical manner, then the ratio
$\omega_{pe}/\Omega_e$ is expected to vary accordingly. If this variation takes place near the upper limit for X1 waves,
say between 0.3 and 0.5, then we expect that the emission of F waves will be on and off spatially when the energetic
electrons pass through these regions. [In regions where  $0.3< \omega_{pe}/\Omega_e<0.4$, the emission is on but in
regions where the plasma has $0.4< \omega_{pe}/\Omega_e<0.5$  the emission is off.] On the other hand, H waves are not
affected because the instability for X2 waves operates over the range $0.4< \omega_{pe}/\Omega_e<0.5$ (or higher). In
this case the pair emission appears to be IIIb-III, and no III-IIIb or IIIb-IIIb can occur. Of course, the same
explanation may also explain the case in which the type IIIb bursts consist only one component. For example, if the ratio
$\omega_{pe}/\Omega_e$ varies around the upper limit for X2 waves, say $\omega_{pe}/\Omega_e\approx 1.4$, then no
emission of X1 waves is possible but emission of X2 waves occurs intermittently in space. As a result we would observe a
single-band type IIIb burst.

We consider that the type IIIb bursts is attributed to the same emission process as type III bursts but in the case the
radiation is modulated by the density and/or magnetic spatial-variation structures along the path that the streaming
electrons pass through. The next question is what would be the cause (or causes) of the spatial structures of the density
and/or magnetic field. Ultra-low frequency (ULF) magnetohydrodynamic (MHD) waves and oscillations, such as magnetosonic
waves, Alfv\'en waves and sausage modes, may be one of the most possible causes, because they exist widely in the solar
atmosphere as being demonstrated comprehensively both from observations and in theories
\citep{Roberts2000,Aschwanden2004,Nakariakov2005,Moortel2012}. In the present article we will not elaborate the theory of
the generation of these waves. We just hypothetically postulate that occasionally there are these waves in the source
region of type IIIb bursts. Our discussion will focus at the consequences of these waves on the fine structures of type
IIIb bursts.

\subsection{Modulation of the frequency ratio $\omega_{pe}/\Omega_e$ }

From the above discussion, one can speculate intuitively that the details of the observed fine structures seen in a
dynamic spectrum of type IIIb burst are determined by the pattern of spatial structures of the ULF waves that modulate
the ratio  $\omega_{pe}/\Omega_e$ in the source region of the radio emission. In general, spatial variation of the wave
field strength as well as the modulated frequency $\omega_{pe}/\Omega_e$  may be different from case to case. For the
purpose of illustration and without loss of generality, in this paper, we just discuss three simple cases, (namely, the
modulation of the ratio $\omega_{pe}/\Omega_e$ by a monochromatic wave, by a standing wave, and by two waves, which
propagate parallel along the ambient magnetic field), to show what kinds of wave form can produce the different fine
structure patterns observed in type IIIb bursts.

\textit{a) Modulation of $\omega_{pe}/\Omega_e$  by a monochromatic wave, single stria bursts}

Let the ambient magnetic field be in the $z$-direction. Note $(\omega_{pe}/\Omega_e)_0$  is the frequency ratio in the
absence of the wave, then the modulated ratio by a monochromatic ULF wave with parallel propagation can be simply modeled
as
\begin{equation}
  \frac{\omega_{pe}}{\Omega_e}=\left(\frac{\omega_{pe}}{\Omega_e}\right)_0[1-\delta_0 \cos(\omega t-k_z z)], \label{1}
\end{equation}
where  $\omega$ and $k_z$  are the wave frequency and the wave number along the ambient magnetic field of the ULF wave,
and $\delta_0$ is the amplitude of the modulation factor. Based on the discussion in the pre-subsection, if
$(\omega_{pe}/\Omega_e)_0$ just happen to be a marginal value  $(\omega_{pe}/\Omega_e)_C$, (e.g. as discussed in Section
3.1), for the excitation of the F waves or H waves so that the emission may be on when
\begin{equation}
  \frac{\omega_{pe}}{\Omega_e}=\left(\frac{\omega_{pe}}{\Omega_e}\right)_C[1-\delta_0],
\end{equation}
and the emission may be off when
\begin{equation}
  \frac{\omega_{pe}}{\Omega_e}=\left(\frac{\omega_{pe}}{\Omega_e}\right)_C[1+\delta_0].
\end{equation}
Where we have made use of the minimum and maximum value of the modulation factor  $g_1(t,z)=\cos(\omega t-k_z z)$.
Obviously, the present scenario works only if  $\delta_0$ is sufficiently large.

If we write $\omega=k_z v_{ph}$  and  $t=z/v_b$, where $v_{ph}=\omega/k_z$ is the wave phase speed along the ambient
magnetic field and $v_b$ is the beam speed of energetic electrons, then the modulation factor seen by the energetic
electrons at different space position is
\begin{equation}
  g_1(t,z)=\cos\left[\left(1-\frac{v_{ph}}{v_b}\right)k_z z\right]\equiv \cos(k_D z) =\cos\left(\frac{2\pi z}{\lambda_D}\right)
\end{equation}
where  $k_D$ is the effective wave number seen by the streaming electrons due to Doppler effect, and  $k_D\equiv
(1-v_{ph}/v_b)k_z$. If we assume that the ECM instability operates when $g_1 >0.7$, we can expect that the type IIIb
bursts would include a number of stria fine structures. In addition, each element of the fine structures has only one
striation in this condition, comparing to the split-pair or triple stria burst which include two or three striations in
each element that will be discussed in the following.

\textit{b)  Modulation of $\omega_{pe}/\Omega_e$  by a standing wave or two waves, double and triple stria bursts.}

First, we study the case that there is a standing wave. Occasionally, low frequency MHD waves may be produced high above
the source region of type IIIb bursts. Since both the ambient magnetic field and density change with altitude, the
Alfv\'en speed varies too. As the wave propagates to a lower altitude, it experiences a higher Alfv\'en speed (so the
wave phase speed). This can be seen from the following relation
\begin{equation}
  v_A=\sqrt{\frac{m_e}{m_p}}\frac{c}{\omega_{pe}/\Omega_e}
\end{equation}
where  $m_e$  and $m_p$  are the mass of electron and proton,  $c$ and $v_A$ are the speed of light and the Alfv\'en
speed. Inside the flux tube, the ratio $\omega_{pe}/\Omega_e$ generally decreases with the decreasing of the altitude, so
the Alfv\'en speed increases with the deceasing of altitude. This means the refractive index in the z direction
decreases. As a result, we consider the descending-propagation MHD wave may be reflected at a low altitude of the source
region of the radio emission. For example, \citet{An89} have demonstrated that transient Alfv\'en waves propagating from
a region with low Alfv\'en speed to a region with higher Alfv\'en speed can be reflected and even trapped in a cavity. On
the other hand, it is commonly believed there are standing waves in a corona loop due to wave reflection at the two
foot-points of the loop \citep{Wang2011,Li2013}. If we assume the descending waves have a narrow frequency spectrum so
that they may be represented by one wave, and in addition, the descending wave continues for sufficiently long time
period, say many ion gyro periods, and finally that there is no dissipation process during the reflection, then the
superposition of the two wave fields will form a standing wave. One can model the modulated frequency ratio by the
standing wave as
\begin{equation}
  \frac{\omega_{pe}}{\Omega_e}=\left(\frac{\omega_{pe}}{\Omega_e}\right)_0
       \{1-\delta_0 [\cos(\omega t+k_z z)+\cos(\omega t-k_z z +\varphi_0)]\}
\end{equation}
where $\omega$ and $k_z$ are the wave frequency and wave number, $\varphi_0$  is the wave phase difference between the
ascending wave and descending wave. An important assumption in obtaining (6) is that the wave field retains more or less
the same form when it returns to the same height.

Similarly, let $\omega=k_zv_{ph}$ and $t=z/v_b$, the modulation factor seen by the energy electrons has the following
form
\begin{equation}
   g_2(t,z)=\cos(k_z' z+\varphi_0/2)\cos(k_z z-\varphi_0/2), \ \ \ k_z^\prime \equiv k_z(v_{ph}/v_b)
\end{equation}
The function  $g_2$ consists of two parts: one is an envelope and the other is a much shorter wavelength wave. For
convenience we define the envelope as E-wave and the latter wave as S-wave. If the S-wave has a wavelength $\lambda_S$
and the E-wave has a wavelength $\lambda_E$, then we define
\begin{equation}
  n\equiv\frac{\lambda_E}{\lambda_S}=\frac{v_b}{v_{ph}}.
\end{equation}
In figure 3 we plot the numerical result of $g_2$  in terms of $k_z^\prime z$. Several values of $n$  are considered. It
is seen that the characteristics change as the ratio $n$  varies. In the case of $n=3,5$ with $\varphi_0=0$ in each of
the interval $\pi/2\le k_z^\prime z \le 3\pi/2$ and $3\pi/2\le k_z^\prime z \le 5\pi/2$, the function $g_2$ has one peak
close to unity. The feature changes as n increases. For example, when $n$ is 8, there may be one or two peaks depending
upon the threshold values of instability. In the case of $n=12$, we see that in the interval $\pi/2\le k_z^\prime z \le
3\pi/2$ or $3\pi/2\le k_z^\prime z \le 5\pi/2$, there are two peaks or three peaks close to 1. The results with
$\varphi_0=\pi/2$ are similar to that of $\varphi_0=0$. We suggest that the twin peaks represent the double stria bursts
while one or three peaks may explain single or triple stria bursts.

When the beam speed is much larger than the wave phase speed, for example the Alfv\'en speed, type IIIb would show double
or triple stria bursts. One the other hand, there will be only single stria bursts. The upper limit of the frequency
ratio $\omega_{pe}/\Omega_e$ for F-waves and H-waves emission are about 0.4 and 1.4, and the corresponding Alfv\'en speed
is about $0.059\ c$ and $0.017\ c$, respectively. This indicates that much higher value of the electron beam speed is
necessary for producing double or triple stria bursts in F component than that for H component, when the modulation by a
standing Alfv\'en-like wave is the cause of type IIIb bursts.

Second, we consider the case that there are two waves propagating upward along the ambient magnetic field. The transient
spatial variation of the modulated ratio can be modeled as
\begin{equation}
  \frac{\omega_{pe}}{\Omega_e}=\left(\frac{\omega_{pe}}{\Omega_e}\right)_0
       \{1-[\delta_0 \cos(\omega_1 t-k_{z1} z)+\delta_0 \cos(\omega_2 t-k_{z2} z +\varphi_0)]\}.
\end{equation}
Here we simply assumed two waves are with the same wave amplitude, and their frequencies and wave numbers are
$(\omega_1,\ k_{z1})$ and $(\omega_2,\ k_{z2})$, respectively.  $\varphi_0$ is the difference between their wave phase
constants. Note
\begin{equation}
 k_1^\prime =\frac{1}{2}\left(1-\frac{v_{ph}}{v_b}\right)(k_{z1}+k_{z2}),
\end{equation}
and
\begin{equation}
 k_2^\prime =\frac{1}{2}\left(1-\frac{v_{ph}}{v_b}\right)(k_{z1}-k_{z2})>0,
\end{equation}
the modulated frequency ratio seen by the streaming electrons is
\begin{equation}
\frac{\omega_{pe}}{\Omega_e}=\left(\frac{\omega_{pe}}{\Omega_e}\right)_0
       \left[1-2\delta_0 \cos(k_1^\prime z+\varphi_0/2)\cos(k_2^\prime z -\varphi_0/2)\right].
\end{equation}
And, the modulation factor is then written as
\begin{equation}
   g_3(t,z)\equiv\cos(k_1^\prime z+\varphi_0/2)\cos(k_2^\prime z -\varphi_0/2)
\end{equation}

The factor $g_3$  has the same function form as that of  $g_2$, which also consists two parts: one is an envelope and the
other is a shorter wavelength wave. Now, the ratio between wavelength of the E-wave and that of the S-wave is defined as
\begin{equation}
   n \equiv \frac{\lambda_E}{\lambda_S}=\frac{k_1^\prime}{k_2^\prime}=\frac{k_{z1}+k_{z2}}{k_{z1}-k_{z2}}
\end{equation}
Thus, one can expect that the modulation of the frequency ratio $\omega_{pe}/\Omega_e$  by two monochromatic waves also
can produce single, double or triple stria bursts depending on the difference of their wave numbers or frequencies.

\section{Discussion}

We will first discuss the mean frequency interval between the adjacent elements in a chain of stria bursts, where an
element may include one striation for single stria burst, two striations for split-pair burst, or three striations for
triple burst. Based on our scenario, this frequency interval is determined by two parameters, namely, the gradient of the
ambient magnetic strength with altitude and the wavelength of MHD waves producing the modulation. In this discussion, the
magnetic field in an active region is modeled by a unipolar magnetic field of a sunspot-field model with the maximum
field strength 2000 G at the center of the spot on the photosphere and the spot radius $0.05\ R_\odot$
\citep{Zheleznyakov70,Yoon2002}. Figure 4 shows the variation of the electron gyro frequency and its harmonic with
altitude in units of solar radius.

The discussion of a chain of single stria bursts is relatively simple, without loss of generality, so we mainly pay
attention to the split-pair burst and triple stria burst in this section. According to our scenario the frequency
interval between the adjacent elements is that between the adjacent peaks of the E-waves. If the E-wave has a parallel
wavelength  $\lambda_E$ at an altitude $H$ where the gyro frequency is $f_g=\Omega_e/2\pi$, then at altitude
$H+\lambda_E$, the gyro frequency changes to $f_g-\Delta f_g$. On the basis on the cyclotron-maser scenario, the
frequency interval between two elements, $\Delta f_E$, should be  $\Delta f_g/2$ (or $\Delta f_g$) for F-waves (or
H-waves). If we know $n$ and if we denote the frequency gap in a double or triple stria burst by  $\Delta f_S$, then we
find $\Delta f_S=2\Delta f_E/n$. Here we remark that in the present theory the gap $\Delta f_S$ is the same for either
double stria or triple stria burst. This finding seems to be consistent with the observational results discussed in
McLean (1985).

Let us consider two cases separately: (i) the single-band type IIIb emission, and (ii) the IIIb-III pair emission.

\textit{i)  Single Component Type IIIb Bursts}

In this case, the waves are assumed to be X-mode harmonic (X2) waves. On the basis of the proposed scenario the observed
frequencies should be $2f_g$. Since the commonly observed type IIIb bursts have frequencies in the range $20 \sim 60$ MHz
\citep{Noe72,Noe75}, the corresponding altitudes are in the range $0.5 \sim 1.0$ solar radii as shown in figure 3. Making
use of these results we can calculate  $\Delta f_S$, which is defined earlier, versus altitude if a wavelength
$\lambda_S$ of the S-wave is assumed. In the present discussion we consider several wavelengths of the S-wave. These are
postulated to be 200, 500, 1000 and 5000 kilometers. The purpose is to see which wavelength would yield the most
reasonable results in comparison with observations. Using the scheme stated in the preceding section we estimate the
frequency gap $\Delta f_S$  in a double or triple stria as a function of altitude. The results are plotted in figure 5a
and figure 5b (solid lines). In figure 5a,  $\Delta f_S$ is expressed as a function of altitude, whereas in figure 5b, is
shown versus the emission frequency. Then we can further calculate  $\Delta f_E$ (for a given value of
$n=\lambda_E/\lambda_S$), the frequency interval between (envelope) elements, which are also shown in figure 5a and
figure 5b (dashed lines) by assuming $n=8$.

From the obtained results we draw the following conclusions:

$\bullet$ The frequency interval $\Delta f_S$  or  $\Delta f_E$ increases with the emission frequency. For example, if we
choose to consider the case in which the MHD wavelength is 1000 km, we find that at emission frequency of 20 MHz the
frequency gap $\Delta f_S$ is 70 kHz while at emission frequency of 60 MHz the frequency gap $\Delta f_S$ is about 350
kHz. The corresponding value of $\Delta f_E$ varies from 0.56 MHz at 20 MHz to 2.9 MHz at 60 MHz.

$\bullet$ According to the present theory, the occurrence of triple stria bursts requires relatively high
$\lambda_E/\lambda_S$ ratio. When the frequency ratio  $\omega_{pe}/\Omega_e$ is modulated by a standing wave, this
happens either with a high beam speed or at high altitudes where the wave phase speeds as well as the radiation
frequencies are low.

\textit{ii) IIIb-III Pair Emission}

We now move on to discuss the F-H pair emission in which under certain conditions the F waves may have structured
emission while the H waves like a normal type III burst. The observed frequencies of the fine structures should be $f_g$.
The altitudes for the emission with frequencies in the range 20 $\sim$ 60 MHz are in the range 0.3 $\sim$ 0.8 solar radii
as shown in figure 4.

Figure 6a and 6b show the variation of the frequency interval $\Delta f_E$  (dashed lines) and the frequency gap $\Delta
f_S$ (solid lines) with the altitude and with the emission frequency, respectively. In this case we find that most of the
conclusions obtained in the above subsection still hold. As the radiation frequency increases, the frequency interval of
both the E-wave and S-wave increases as well. It is found that, in the case in which the wavelength $\lambda_S$  is 1000
km, the frequency interval $\Delta f_S$ (or $\Delta f_E$) at emission frequency 20 MHz and 60 MHz are about 100 kHz (or
0.6 MHz) and 500 kHz (or 3.1 MHz), respectively. The frequency intervals are slightly larger than that for a single
component type IIIb bursts in section 4.1, because the altitude of the true source region of F wave is lower than that of
the H wave with the same emission frequency, and the gradient of the magnetic field strength increases with the decrease
of altitude.

In addition, observation shows that the fine structures occur most at low frequency or in other word more often at high
altitude. A qualitative discussion for this observation is as follows. According to the present scenario, two necessary
conditions are needed to produce fine structure. First, the ambient frequency ratio $(\omega_{pe}/\Omega_e)_0$  without
MHD waves is near its upper limit for the exciting of ECM instability. This limit is approached more often at high
altitude since the ratio  $(\omega_{pe}/\Omega_e)_0$ is generally increasing with the increase of altitude in the true
source region of the emission in the flux tube. Second, the amplitude of the MHD waves is large enough to modulate the
frequency ratio $\omega_{pe}/\Omega_e$  in a wide range value, so that the radio emission can be switched on and off
spatially. Since both the ambient magnetic field strength and the plasma density decrease with the increase of altitude,
their values can be disturbed more easily at high altitude than at lower altitude. For example, assuming an Alfv\'en wave
propagating upward along the flux tube from an altitude $H_1$  to the altitude $H_2$ without dissipation ($H_2>H_1$), it
is easy to demonstrate from the conservation of wave energy that
\begin{equation}
  \frac{\delta B_2^2 / B_{0,2}^2}{\delta B_1^2 / B_{0,1}^2}=\frac{v_{A1}}{v_{A2}}\frac{B_{0,1}}{B_{0,2}}
\end{equation}
where $\delta B_i$, $B_{0,i}$, and $v_{Ai}$ are the wave amplitude, the ambient magnetic field strength, and the Alfv\'en
speed, respectively. Index "$i=1,2$ " represent parameters at the altitude  $H_1$ and $H_2$. Both the ambient magnetic
field strength and the Alfv\'en speed are generally decreasing the increase of altitude in the flux tube. Thus, the
higher the altitude is, the larger the relative perturbation of the magnetic field strength $\delta B^2/B_0^2$  is
expected to be.

\section{Conclusions}

To summarize, in this paper we propose a possible scenario for the fine structures of type IIIb bursts based on the ECM
emission theory of solar type III radio bursts. The essence of the scenario is that the presence of enhanced MHD waves in
the source region of radio emission, the normal type III emission is modulated by spatial structure of density and/or
magnetic field associated with the ULF MHD waves. The details are discussed in Section 3.

The conclusions drawn from the present theory are consistent with observations. Among them, the following points deserve
attention.

$\bullet$ The proposed scenario is able to resolve the issue why in the events of F-H pair emission only IIIb-III but no
IIIb-IIIb or III-IIIb cases are observed.

$\bullet$ The scenario can explain the occurrence of split pair (or double stria) bursts and triple stria bursts.
Moreover, the scenario also predicts that in a triple stria burst the fine structure is symmetric with respect to the
middle element.

$\bullet$ In general fine structure components of type IIIb emission occur in high altitudes where the radiation
frequencies are low. The frequency-separation of striae in a given burst increases with increasing frequency
\citep{Ellis69,Noe75}.

$\bullet$ Concerning the polarization of type IIIb bursts, X-mode is assumed in the above discussion. However, the
scenario is also applicable for O-mode, since there are upper limit of the frequency ratio $\omega_{pe}/\Omega_e$  for
the emission of O-mode. That is the fundamental O1-mode can be excited in the condition $0.1<\omega_{pe}/\Omega_e<1.0$,
while the condition for the harmonic O2-mode is $0.1<\omega_{pe}/\Omega_e<2.0$ which is much broader. One of the
essential conclusions, which is in agreement with observations as reported in \citet{Ellis69}, \citet{Noe72}, and
\citet{Noe75}, is that the elements of a pair or a triplet stria burst are always polarized in the same sense.

For illustration purpose we present figure 7 in which we depict a calculated dynamic spectrum for the case of a single
component type IIIb emission. This figure is calculated based on the scheme used in \citet{Wu2002}, the wavelength of the
MHD wave is assumed to be 1000 km. In this figure we show a chain of split-pairs and triple stria bursts.

The preliminary results presented in this paper are encouraging for the ECM scenario of type III bursts, but more
discussion is still needed in the future. For example, observation of the type III burst polarization in the literature
seems to favor the O mode, although there are uncertainties (the difficulty stems from the fact that there is no
directional measurement of the polarity of the magnetic field in the corona source region where the radiation come from.)
About this point, the recent work by \citet{Wu2012} deserve attention. It is found that the growth rate of O-mode in the
ECM instability may be significantly influenced by intrinsic Alfv\'en turbulence in the solar corona through modifying
the resonant wave-particle interaction \citep{Wu2012,Wu2014}. And further more, \citet{Zhao2013} applied this theory to
explain the fine structures of type IIIb bursts by assuming the modulation is caused by a linearly polarized
monochromatic Alfv\'en wave. The numerical results show that these effects mainly enhance the growth rate of the
fundamental O1-mode. This seems to favor that the F component is O-mode while the H component is X mode. However,
observations indicate that the polarization of type IIIb bursts are not differ significantly from that of normal F-H
pairs, and the sense of polarization of F and H radiation is invariably the same \citep{Dulk80}. Thus, we do not consider
the influence of the ULF waves on the wave-particle resonant interaction in the present study. While we also suggest the
fine structures are according to intermittent emission of radio waves modulated by ULF MHD waves, the physical process
for the modulation is basically different from that discussed in the paper by \citet{Zhao2013}.

The scenario may be also help us to understand other types of
solar radio burst with fine structures such as the zebra bursts
\citep{Kuijpers80,Ning2000,ChenB2011,YuSJ2013}, which will be
discussed in the future. Finally, as pointed out by the referee,
the main point of the present paper, generation of type IIIb fine
structure in terms of emission modulated by standing (or slowing
traveling) waves in the magnetic field, does not depend on a
specific emission mechanism.







\acknowledgments

The research was supported by the National Science Foundation of China grants 41174123 and 41421063, the Chinese Academy
of Sciences grant KZCX2-YW-QN512 and KZZD-EW-01, and the Fundamental Research Funds for the Central Universities under
grant WK2080000077.

\begin{figure}
\epsscale{0.8} \plotone{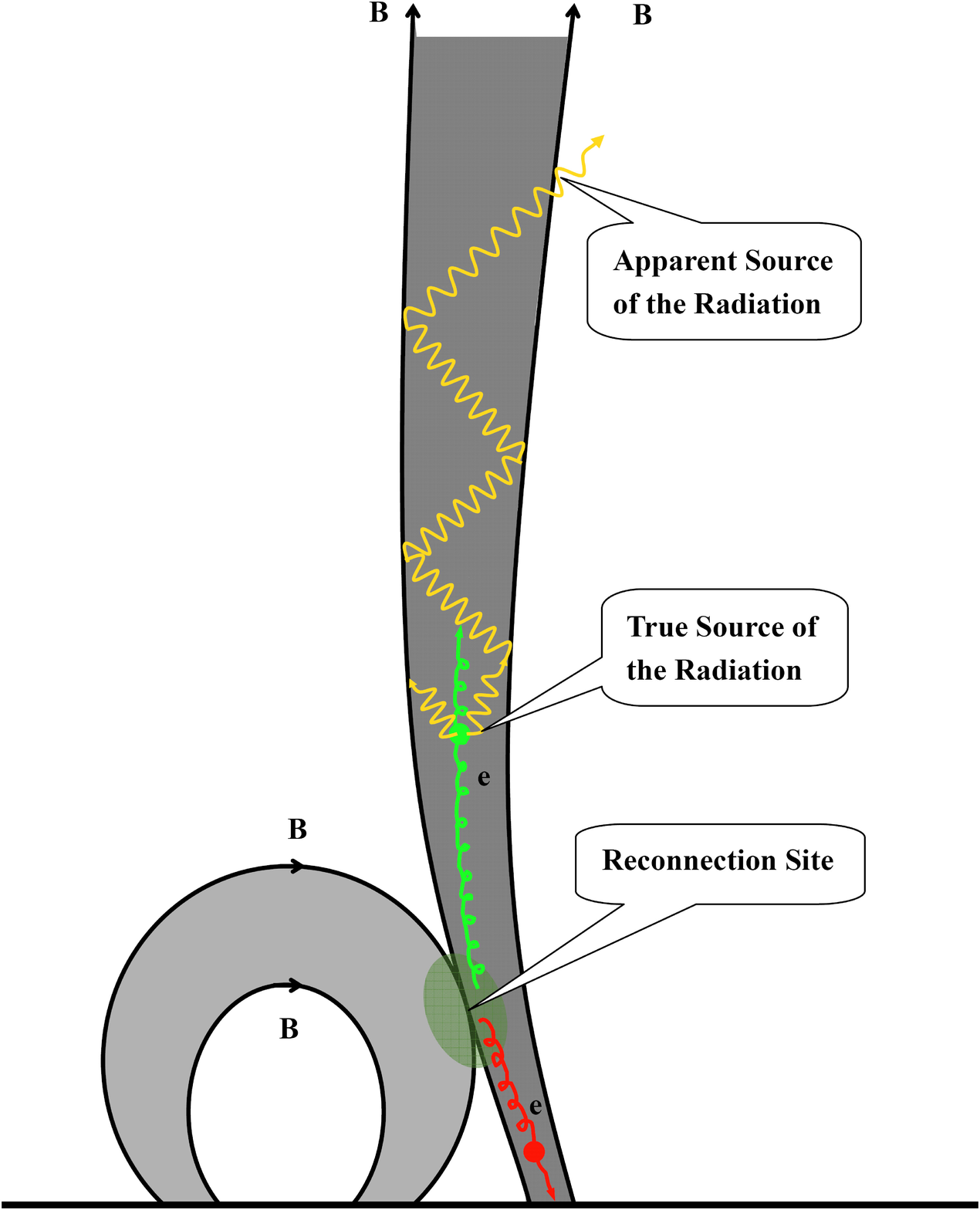} \caption{A graphic description summarizes a scenario of type III bursts based on the
electron cyclotron maser emission mechanism.\label{Fig1}}
\end{figure}


\begin{figure}
\epsscale{.80} \plotone{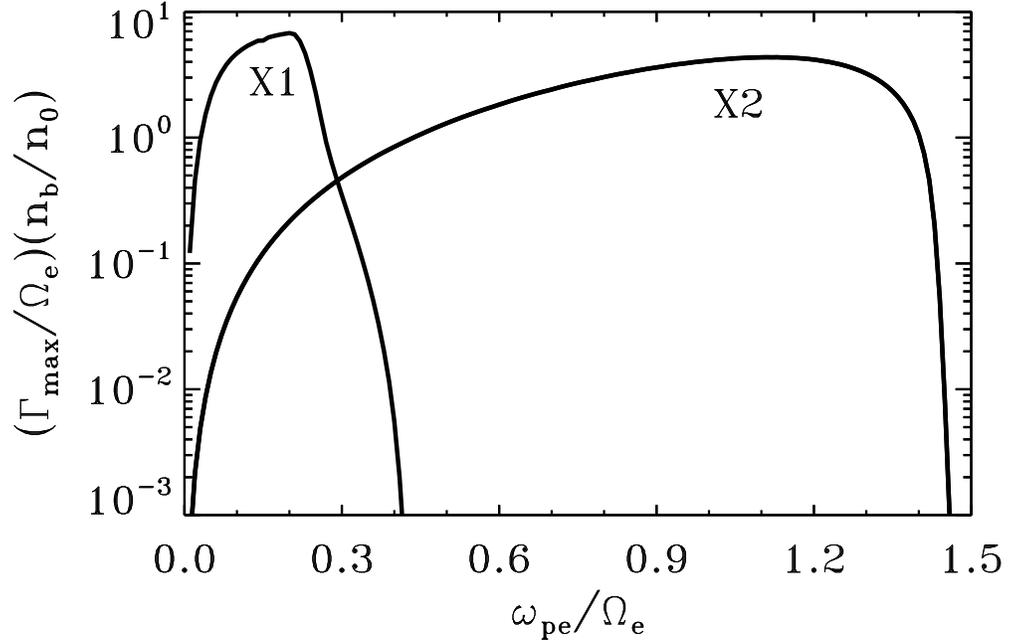} \caption{Maximum growth rates of F (fundamental) waves (X1) and H (harmonic) waves (X2)
are plotted versus the ratio $\omega_{pe}/\Omega_e$, where $n_b$ and $n_0$ are the number density of the energy electrons
and ambient electrons, respectively. The velocity distribution of the energy electrons is modeled as
$F_b(u_\perp,u_\parallel)=C \exp[-(u_\perp -u_{\perp 0})^2/\alpha^2 - (u_\parallel-u_{\parallel 0})^2/\beta^2]$ with
$u_{\perp 0}=u_0\sin\theta$, $u_{\parallel 0}=u_0\cos\theta$, $\alpha=0.1u_{\perp 0}$, $\beta=0.2u_{\parallel 0}$,
$u_0=0.3 c$ and $\theta=20^\circ$, and where $c$ is the speed of light. It is seen that amplification of F waves are restricted to a small range of
$\omega_{pe}/\Omega_e$ only.\label{Fig2}}
\end{figure}



\begin{figure}
\plottwo{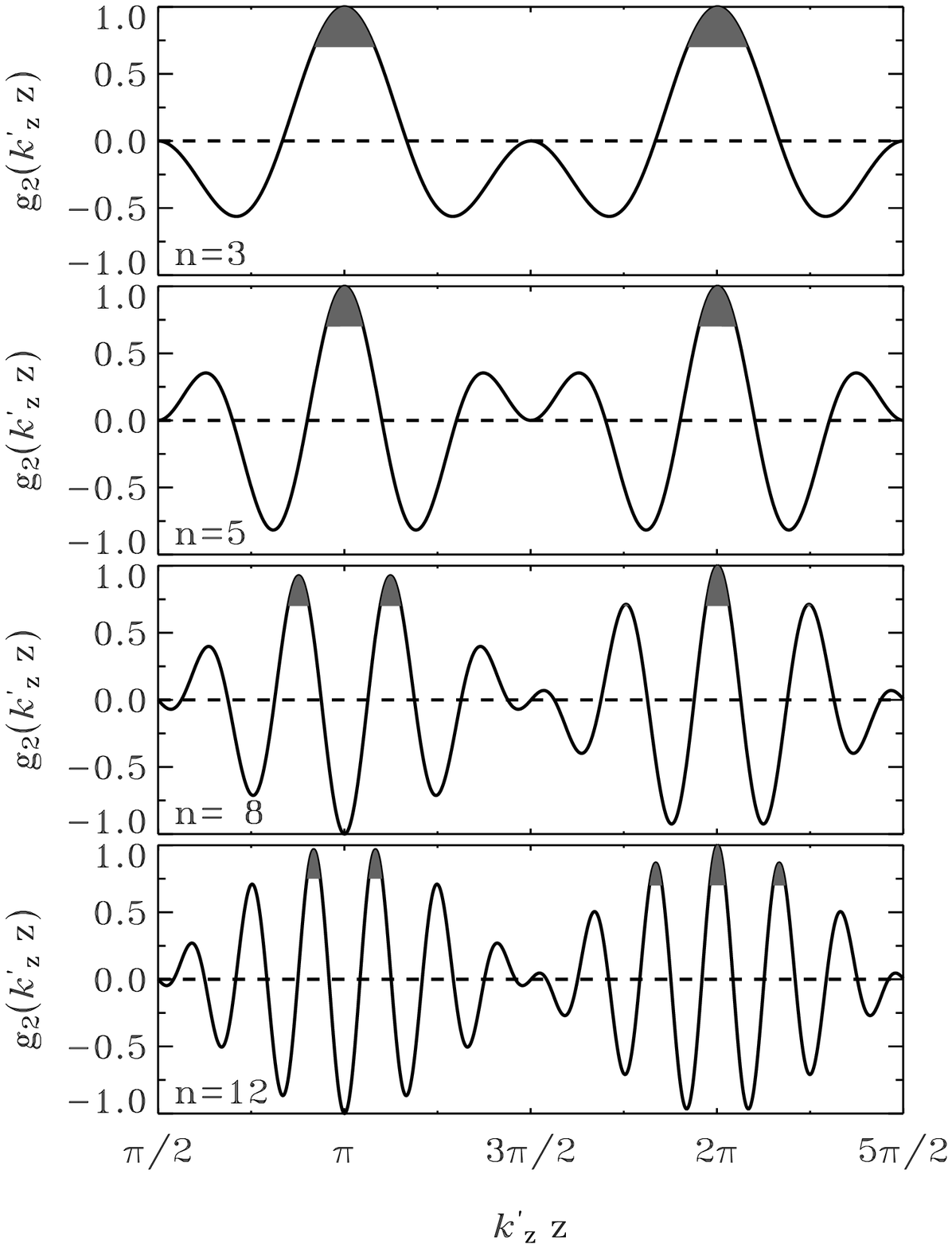}{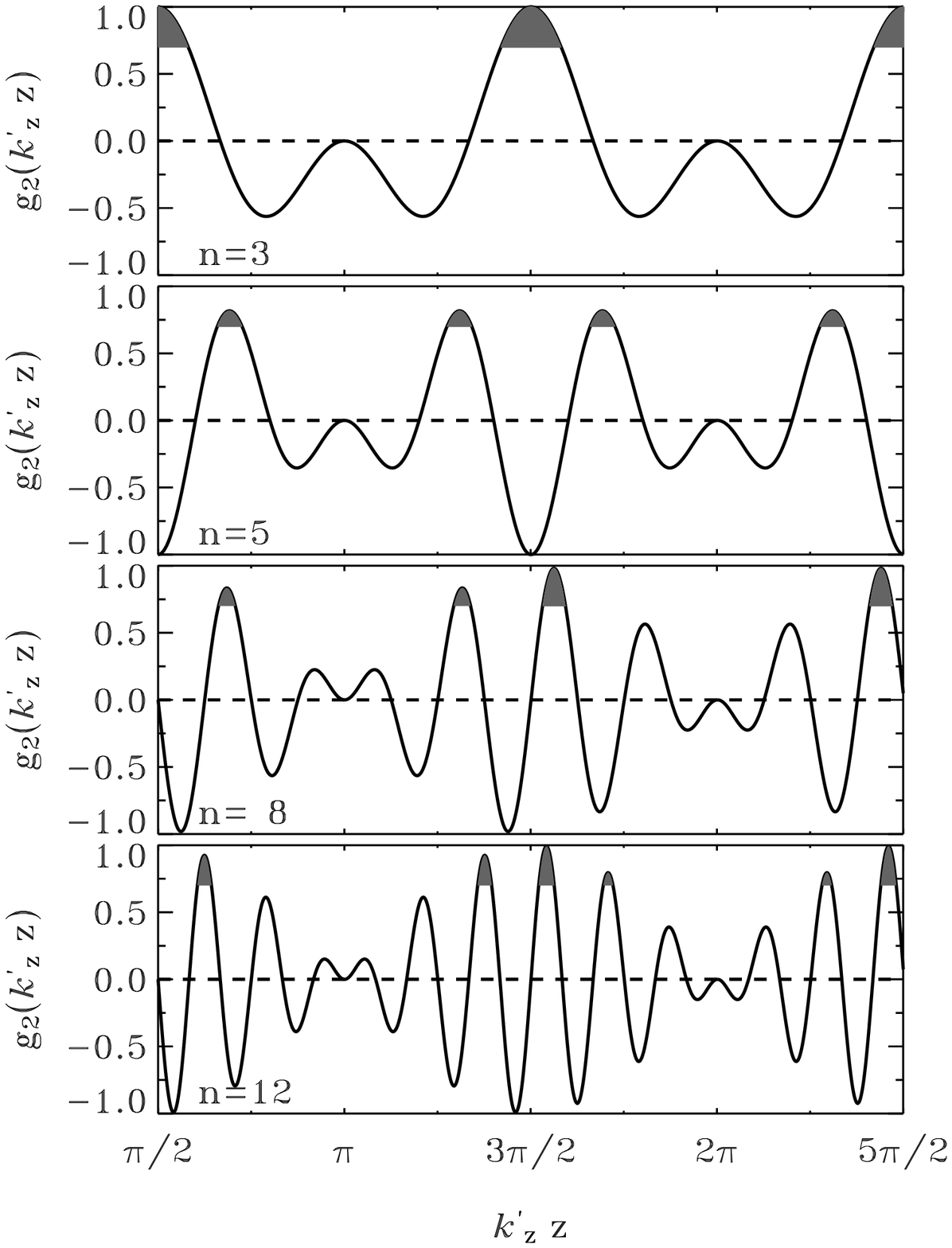} \caption{Modulation factor of the
standing wave is plotted as a function of $k_z^\prime z$ for
several values of  $n$. The left and right panel are for the
results with the wave phase constant $\varphi_0=0$ and
$\varphi_0=\pi/2$, respectively. For illustration purpose, it is
assumed that the maser instability operates when $g_2>0.7$. It is
seen that in the case  $n=3$ or 5, single stria elements may
occur, whereas in the case of $n=8$ and 12 double and/or triple
stria bursts may happen.\label{Fig3}}
\end{figure}


\begin{figure}
\plotone{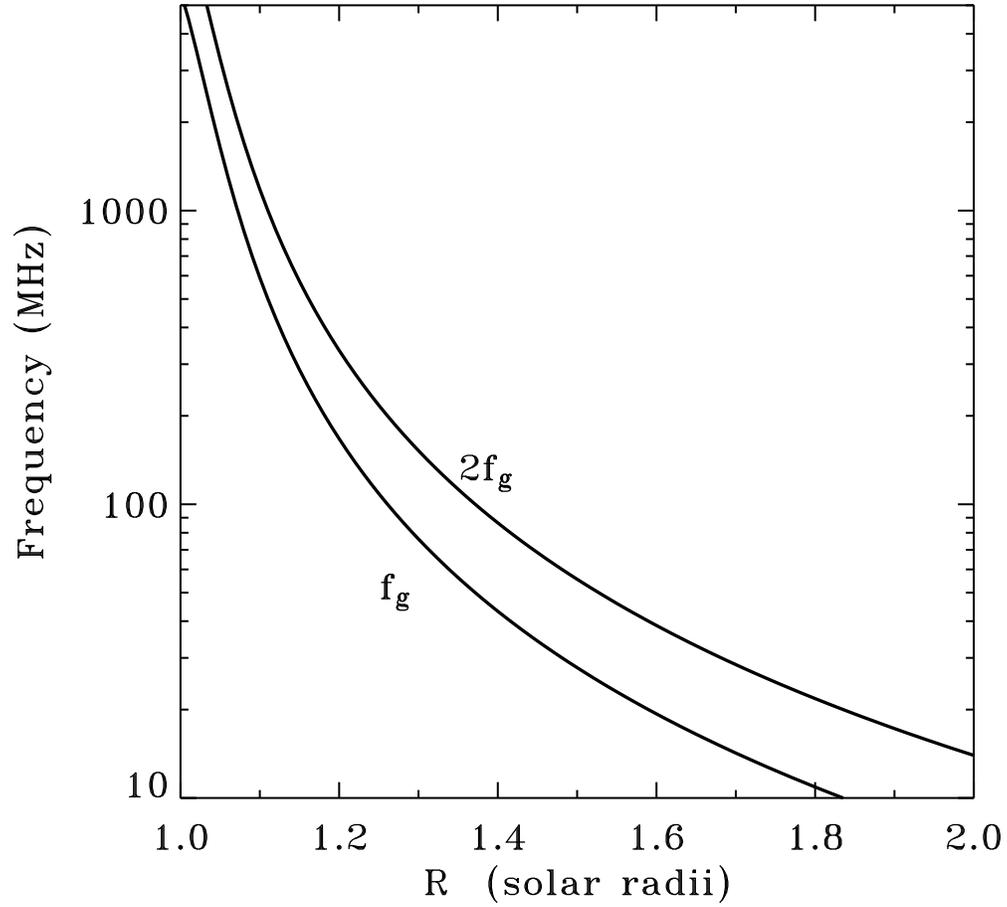} \caption{Variation of the electron gyro-frequency and it harmonics with altitude in the
density-depleted tube, where the magnetic field is modeled by a unipolar magnetic field of a sunspot-field model with the
maximum field strength 2000 Gauss at the center of the spot on  the photosphere and the spot radius 0.05
$R_\odot$.\label{Fig4}}
\end{figure}


\begin{figure}
\epsscale{1.0} \plottwo{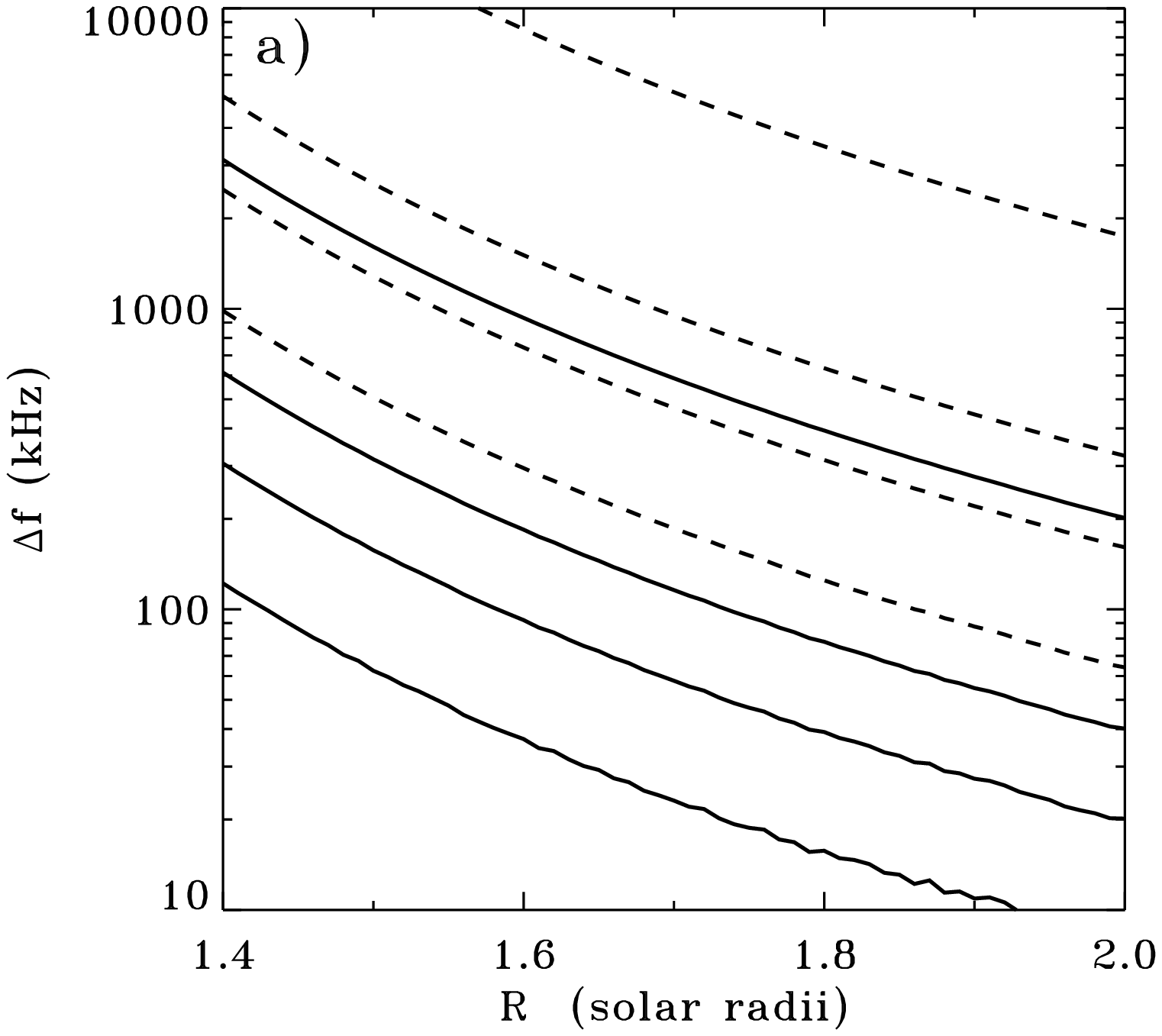}{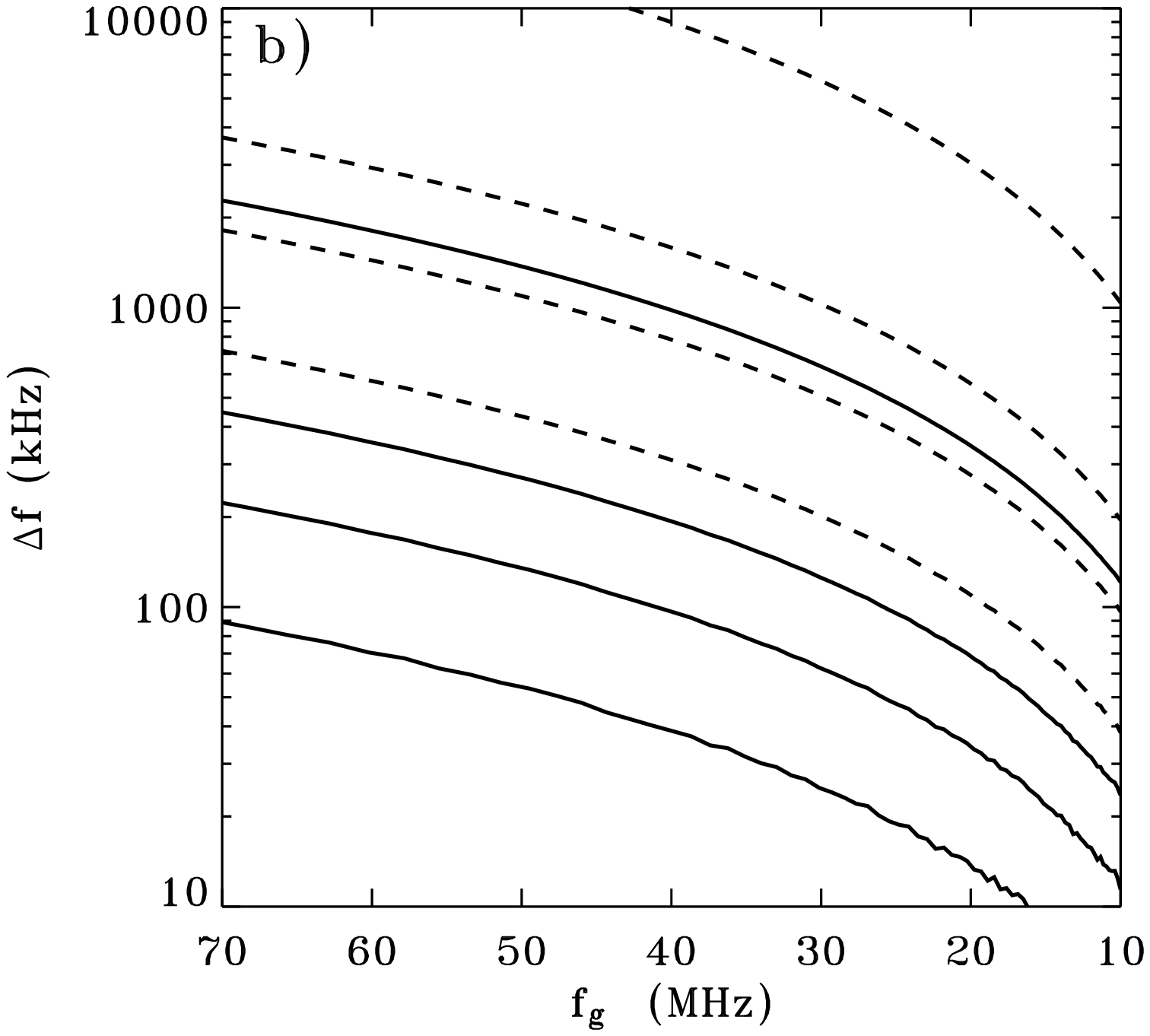} \caption{a) Numerical results of frequency interval $\Delta f$  between
elements and striations for the case of single component type IIIb bursts are shown as function of altitude. The solid
curves represent the interval $\Delta f_S$ in the fine structure while the dash curves are $\Delta f_E$  for the envelope
wave. Several wavelengths of the S-wave are considered. The wavelength of the E-wave is assumed to be 8 times of the
value of the S-wave. Curves from the lower-left corner to the upper-right corner represent the results with wavelength
$\lambda_S=200$, 500, 1000 and 5000 km, respectively. b) The same results presented in (a) are shown as functions of
emission frequency. This is done on the basis of the cyclotron maser scenario from which the altitude is converted to
radiation frequency.\label{Fig5}}
\end{figure}


\begin{figure}
\epsscale{1.0} \plottwo{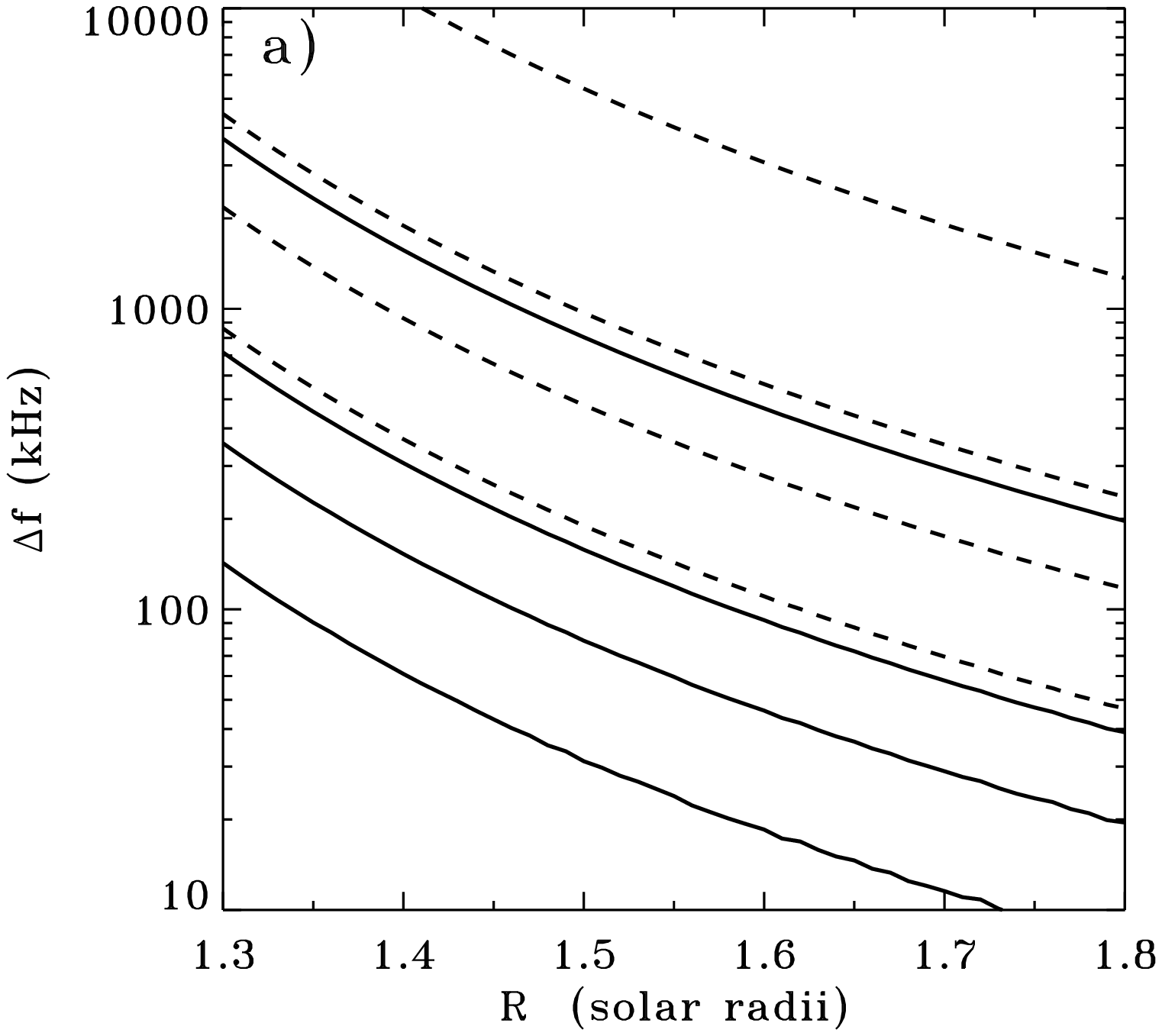}{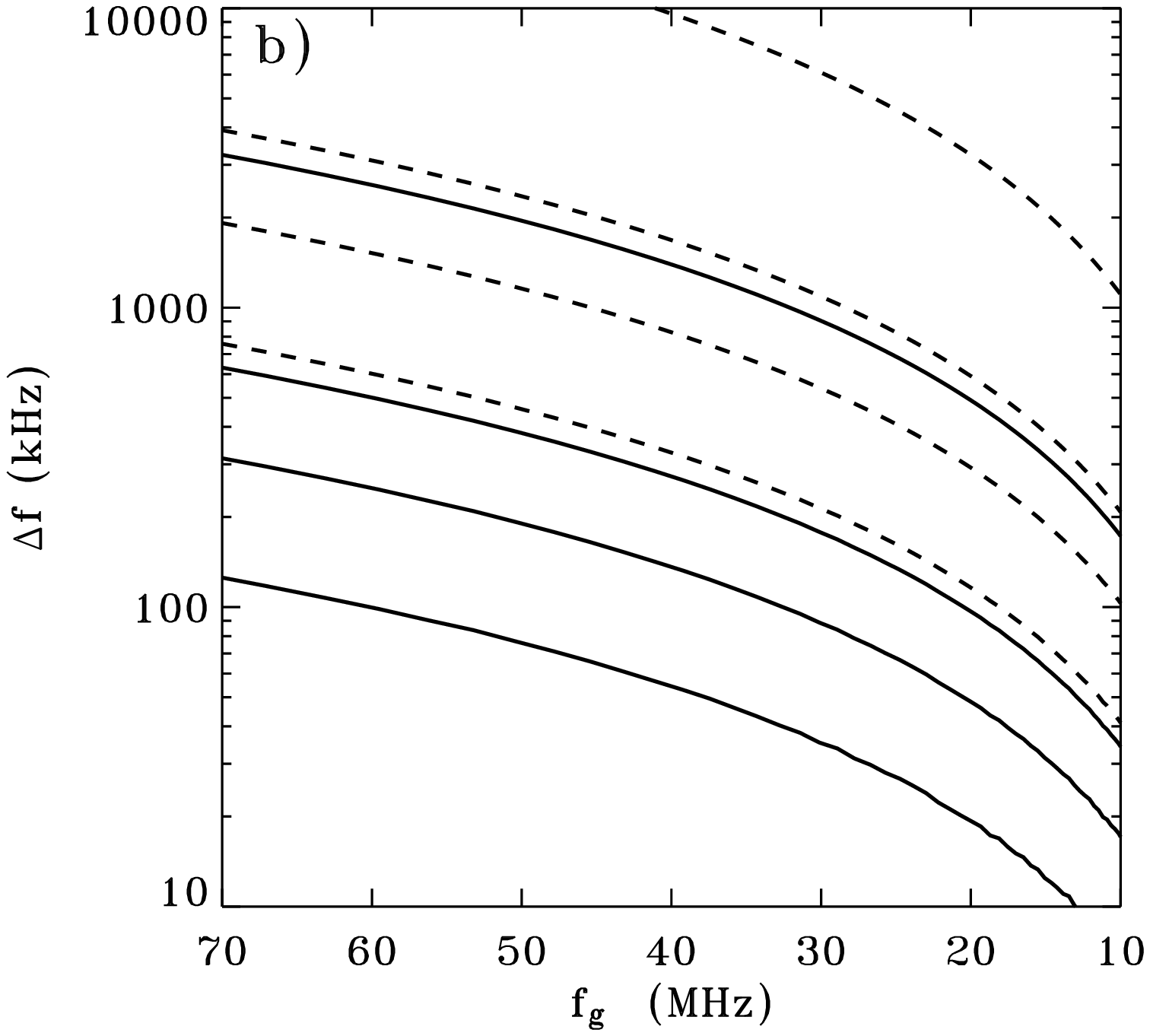} \caption{a) Numerical results of frequency interval $\Delta f$  between
elements and striations for the case of IIIb-III case are shown as function of altitude. The solid curves represent the
interval $\Delta f_S$  in the fine structure while the dash curves are $\Delta f_E$  for the envelope wave. Several
wavelengths of the S-wave are considered. The wavelength of the E-wave is assumed to be 8 times of the value of  the
S-wave. Curves from the lower-left corner to the upper-right corner represent the results with wavelength
$\lambda_S=200$, 500, 1000 and 5000 km, respectively. b) The same results presented in (a) are shown as functions of
emission frequency. This is also done as in figure 5b by converting the altitude to radiation frequency.\label{Fig6}}
\end{figure}


\begin{figure}
\plotone{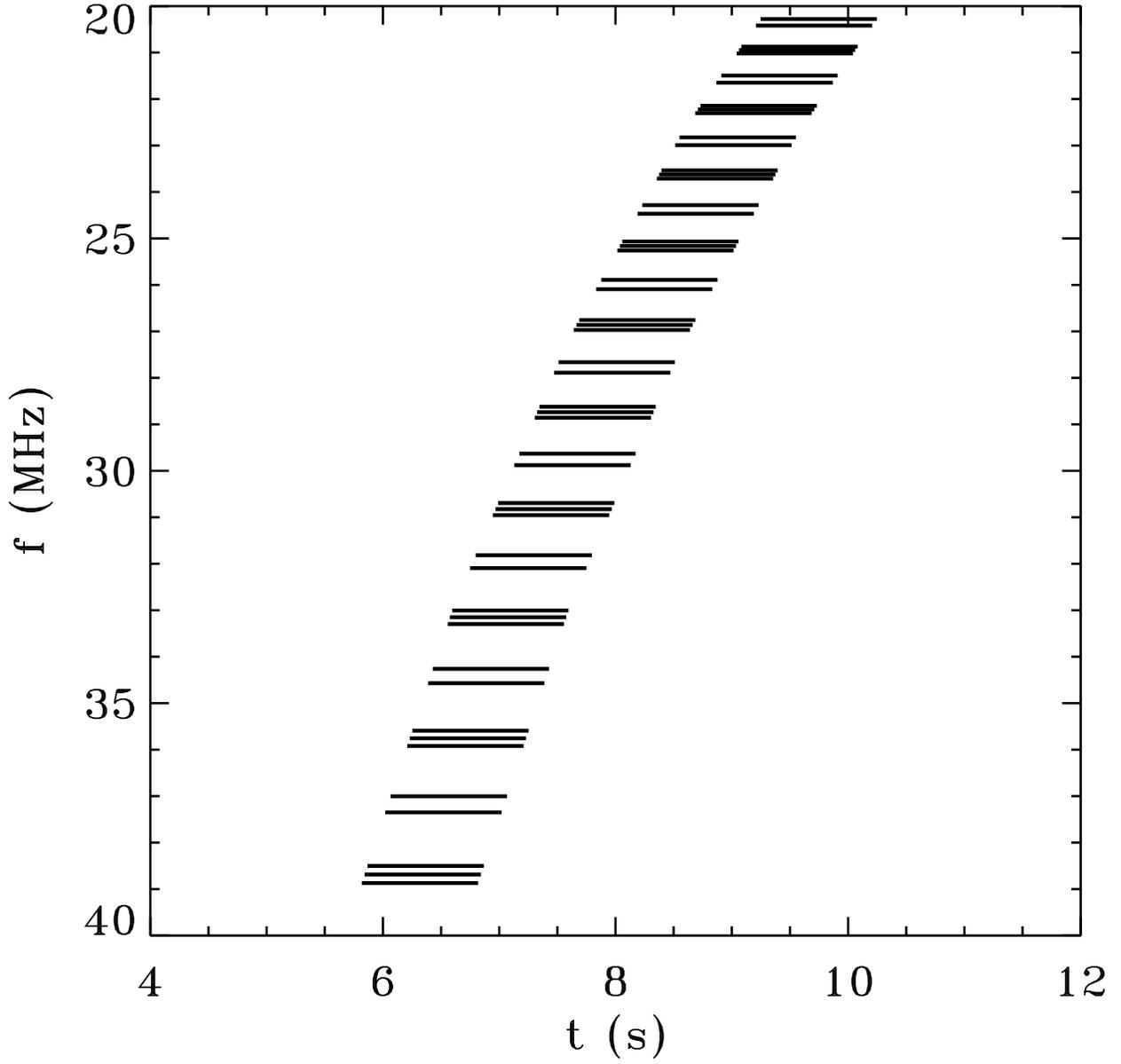} \caption{A numerically created dynamic spectrum for the case of a single component type IIIb emission.
Elements of fine structures with split-pair and triple stria bursts are illustrated. The time duration is arbitrarily
taken to be one second.\label{Fig7}}
\end{figure}




\end{document}